\documentclass[prd,aps,floats,preprint]{revtex4}
\usepackage{graphicx}
\usepackage[english]{babel} % grammar package
\usepackage{braket}%Dirac notation
\newcommand{\be}{\begin{eqnarray}}
\newcommand{\ee}{\end{eqnarray}}
\newcommand{\ba}{\begin{eqnarray}}
\newcommand{\ea}{\end{eqnarray}}

\def\ba{\begin{eqnarray}}
\def\ea{\end{eqnarray}}
\def\be{\begin{equation}}
\def\ee{\end{equation}}

\def\eps{\epsilon}

\def\R{{\mathcal R}}

\newcommand{\bea}{\begin{eqnarray}}
\newcommand{\eea}{\end{eqnarray}}

\newcommand{\eqn}[1]{eq.~(\ref{#1})}
\def\R{{\mathcal R}}
\def\d{{\rm d}}

\usepackage{morefloats} %When there are many figs.~
\usepackage[titletoc,title]{appendix}
\usepackage{color}
\begin{document}

\title{Primordial black holes from local  features of the inflaton potential}

\author{Alexander Gallego Cadavid}
\affiliation{Universidad  de Valpara\'{\i}so,  Avenida Gran Breta\~na 1111,  Valpara\'{\i}so  2360102,  Chile}
\affiliation{Instituto de F\'isica, Universidad de Antioquia, A.A.1226, Medell\'in, Colombia}
\author{Antonio Enea Romano}
\affiliation{Instituto de F\'isica, Universidad de Antioquia, A.A.1226, Medell\'in, Colombia}
\affiliation{ICRANet, Piazza della Repubblica 10, I--65122 Pescara, Italy}

\begin{abstract}
Primordial black holes (PBH) can form in the  early Universe from peaks of the primordial curvature perturbations. The statistics of the peaks determines the abundance of PBHs, and is related to the amplitude of the primordial curvature spectrum. We study single field inflationary models with local features of the inflaton potential which induce a growth of curvature perturbations on super-horizon scales, which can increase the spectrum, and consequently the predicted PBHs abundance.  
We compute the effects of the different parameters of the features on the PBHs abundance, giving some example of models compatible with observational constraints. The mechanism is general, and can induce the PBH production in diffent mass ranges by appropriately tuning the parameters, and we give a specific example producing PBHs with abundance compatible with asteroid mass constraints.

\end{abstract}

\email{antonioenea.romano@ligo.org}
\maketitle

\section{Introduction}
In the standard cosmological model primordial curvature perturbations are the seeds of the  large scale structure  of the Universe.
When these perturbations are sufficiently large \cite{Sasaki:2018dmp} primordial black holes (PBH) are formed, which could have important observational consequences \cite{Sasaki:2016jop,Josan:2009qn,Carr:2009jm,Belotsky:2014kca,Belotsky:2018wph, Green_2021, Carr:2020gox}.
In this paper we investigate the PBH formation in single field models with local features in the inflation potential.
In the slow-roll regime curvature perturbation is freezing on super-horizon scale, but a temporary violation of the slow-roll regime due to feature in the potential can induce a super-horizon growth which can enhance the curvature spectrum around the scales affected by the super-horizon curvature growth.
The effects of these features were studied in the past with focus on non-gaussianity \cite{GallegoCadavid:2015hld}, and as possible explanations to the cosmic microwave background (CMB) anomalies \cite{GallegoCadavid:2016wcz}, but as mentioned above the slow-roll violation associated to these features is also the main primordial curvature perturbation spectrum enhancement  mechanism  \cite{Romano:2016gop}, justifying the expectation of increased PBH production in these models. 

While in  slow-roll inflationary models $\R$ is expected to be conserved on super-horizon scales \cite{Wands:2000dp}, if slow-roll is violated  there can be a super-horizon evolution of $\R$, such as in globally adiabatic models \cite{Romano:2016gop} or inflation with a quasi-inflection point in the potential \cite{Ezquiaga:2018gbw}. 
The super-horizon growth of $\R$ has profound implications because it can give rise to PBH production  \cite{Garcia-Bellido:2017mdw}, with different important observable effects.
We study the affects of the different feature parameters on the PBH abundance, and confirm that there is a high sensitivity, as pointed out in Ref.~\cite{Cole:2023wyx}.
By tuning the parameters of the feature it is possible to produce PBHs in different mass ranges, and we consider a specific example of a model producing PBHs with abundance compatible with current asteroid mass constraints.

\section{Single field slow-roll inflation}
We consider inflationary models with a single scalar field and a standard kinetic term according to the action \cite{Liddle:2000cg}
\begin{equation}\label{action1}
  S = \int d^4x \sqrt{-g} \left[ \frac{1}{2} M^2_{Pl} R  - \frac{1}{2}  g^{\mu \nu} \partial_\mu \phi \partial_\nu \phi -V(\phi)
\right],
\end{equation}
where $ M_{Pl} = (8\pi G)^{-1/2}$ is the reduced Planck mass,
$g_{\mu \nu}$ the flat Friedmann-Lemaître-Robertson-Walker (FLRW) metric, $R$ the Ricci scalar, and $V$ is the potential
energy of the inflaton field $\phi$. The variation of the action with respect to the metric
and the scalar field gives the Friedmann equation and the equation of motion of the inflaton
\begin{equation}\label{ema}
  H^2 \equiv \left(\frac{\dot a}{a}\right)^2= \frac{1}{3 M^2_{Pl}}\left( \frac{1}{2} \dot \phi^2 + V(\phi) \right),
\end{equation}
\begin{equation}\label{emphi}
  \ddot \phi + 3H\dot \phi + \partial_{\phi}V = 0,
\end{equation}
where $H$ is the Hubble parameter and we denote with dots and $\partial_{\phi}$ the derivatives with respect to time and scalar field, respectively. 
The definitions we use for the slow-roll parameters are
\bea \label{slowroll}
  \epsilon \equiv -\frac{\dot H}{H^2} \,\, \mbox{ and } \,\, \eta \equiv \frac{\dot \epsilon}{\epsilon H}.
\eea

In order to study the production and abundance of PBH we will consider a potential energy given by \cite{GallegoCadavid:2015hld, GallegoCadavid:2016wcz}
\be\label{pot}
V(\phi)= V_{0}(\phi) + V_{\rm LF}(\phi) \, ,
\ee
where $V_{0}(\phi)$ is a featureless potential and $V_{\rm LF}$ corresponds to a local feature (LF), first studied in Ref.~\cite{GallegoCadavid:2015hld}. For simplicity we will consider the case of a linear inflaton potential
\bea
V_0(\phi)=\gamma \phi \,.
\eea
In order to fulfill the CMB COBE/Planck  normalization \cite{Ade:2015lrj,Planck:2013jfk},  we chose as initial conditions $\gamma=3.2767\times10^{-10}, a(t_i)=1, \phi(t_i)=9.81369$ and $\dot \phi(t_i)=-3.05691\times10^{-6}$, where $t_i=0$ is the initial time of the integration and all units are given in Planck mass. From now on we adopt a system of units in which $c=\hbar=M_{Pl}=1$.
%**************************************************
%**************************************************
\begin{figure}
  \includegraphics[scale=1]{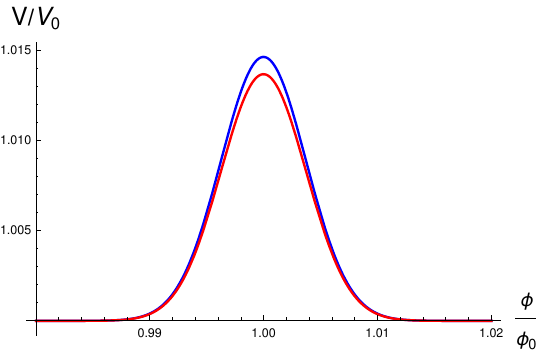}
 \caption{$V/V_0$ is plotted for $\sigma=0.02$, $n=1$,  $\lambda=1.8196427\times 10^{-11}$ (blue) and $\lambda=1.7\times 10^{-11}$ (red).}
\label{vlambdaback}
\end{figure}

\begin{figure}
 \begin{minipage}{.45\textwidth}
  \includegraphics[scale=0.8]{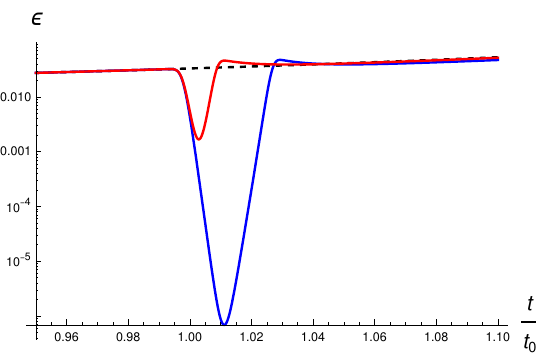}
  \end{minipage}
 \begin{minipage}{.45\textwidth}
  \includegraphics[scale=0.8]{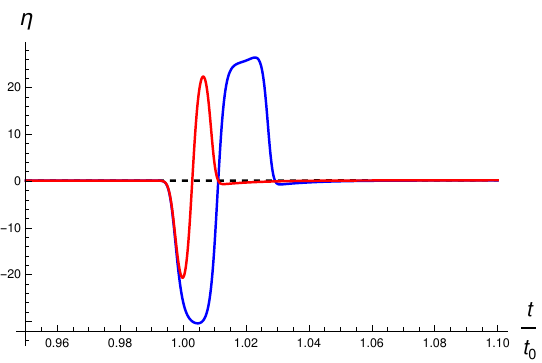}
 \end{minipage}
 \caption{ The slow-roll parameters $\epsilon$ and $\eta$ are plotted for $\sigma=0.02$, $n=1$,  $\lambda=1.8196427\times 10^{-11}$ (blue) and $\lambda=1.7\times 10^{-11}$ (red). The dashed black lines correspond to the featureless behavior.}
\label{lambdaback}
\end{figure}
%**************************************************
\begin{figure}
  \includegraphics[scale=1]{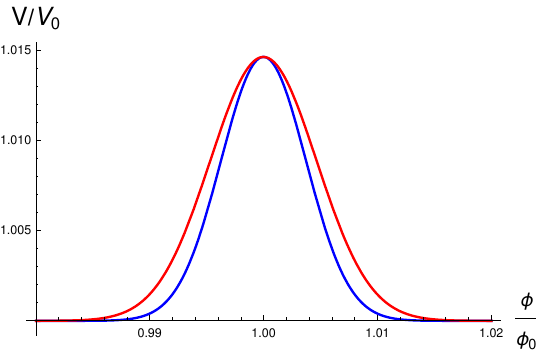}
  \caption{$V/V_0$ is plotted for $\lambda=1.8196427\times 10^{-11}$,  $n=1$, $\sigma=0.02$ (blue) and $\sigma=0.025$ (red).}
\label{vsigmaback}
\end{figure}

\begin{figure}
 \begin{minipage}{.45\textwidth}
  \includegraphics[scale=0.8]{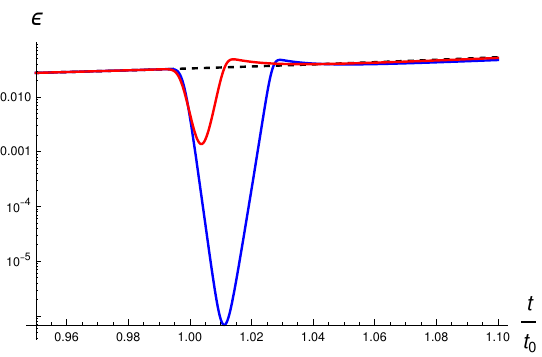}
  \end{minipage}
 \begin{minipage}{.45\textwidth}
  \includegraphics[scale=0.8]{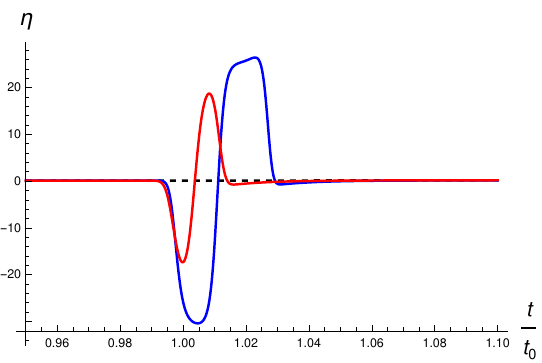}
 \end{minipage}
 \caption{The slow-roll parameters $\epsilon$ and $\eta$ are plotted for $\lambda=1.8196427\times 10^{-11}$,  $n=1$,  $\sigma=0.02$ (blue) and $\sigma=0.025$ (red). The dashed black lines correspond to the featureless behavior.}
\label{sigmaback}
\end{figure}
%**************************************************
\begin{figure}
  \includegraphics[scale=1]{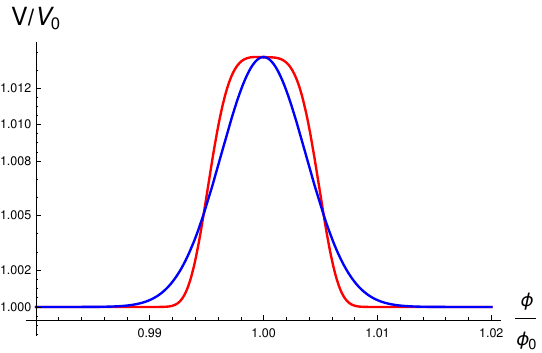}
  \caption{$V/V_0$ is plotted for $\lambda=1.707869\times10^{-11}$, $\sigma=0.02$,  $n=2$ (blue) and $n=4$ (red).}
\label{vnback}
\end{figure}

\begin{figure}
 \begin{minipage}{.45\textwidth}
  \includegraphics[scale=0.8]{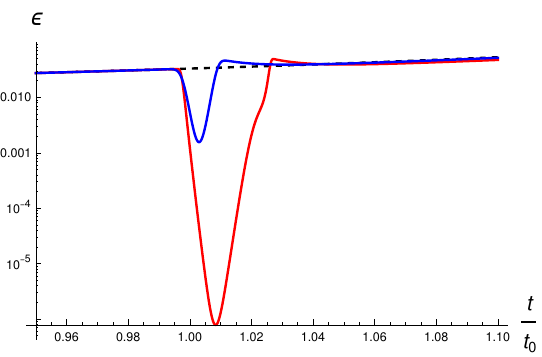}
  \end{minipage}
 \begin{minipage}{.45\textwidth}
  \includegraphics[scale=0.8]{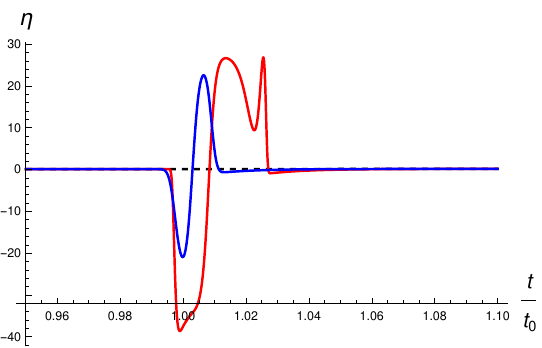}
 \end{minipage}
 \caption{The slow-roll parameters $\epsilon$ and $\eta$ are plotted for $\lambda=1.707869\times10^{-11}$, $\sigma=0.02$,  $n=2$ (blue) and $n=4$ (red). The dashed black lines correspond to the featureless behavior.}
\label{nback}
\end{figure}
%**************************************************
%**************************************************

As for the feature, it is characterized by a step symmetrically dumped by an even power negative exponential factor \cite{GallegoCadavid:2015hld, GallegoCadavid:2016wcz}
\be
V_{\rm LF}(\phi)=\lambda e^{-( \frac{\phi-\phi_0}{\sigma})^{2 n}} \, , \label{LF}
\ee
which is symmetric with respect to the location of the feature and it is only affected in a limited range of the scalar field value. Due to this the spectrum is only modified in a narrow range of scales. 

The effects of the feature on the featureless linear potential and the slow-roll parameters are shown in Figs. \ref{vlambdaback}-\ref{nback}. We can see that there are oscillations of the slow-roll parameters around the feature time $t_0$, defined as $\phi_0=\phi(t_0)$ \cite{GallegoCadavid:2015hld}. The magnitude  of the potential modification is controlled by the parameter $\lambda$, as its effect is such that larger value of $\lambda$ give larger values of the potential around $\phi_0$ and the slow-roll parameters, see Figs. \ref{vlambdaback}-\ref{lambdaback}. This is due to a larger feature of the potential which induces a larger time derivative of the Hubble parameter. In Figs. \ref{vsigmaback}-\ref{sigmaback} we can see that the size of the range of field  values where the potential is affected by the feature is determined by the parameter $\sigma$ such that the slow-roll parameters are smaller for larger $\sigma$, since a larger width of the feature tend to reduce the time derivative of the Hubble parameter. In this case, in fact, the modification of the potential is also associated to smaller derivatives with respect to the field, since the shape of the potential is less steep. An interesting case is that of the effect of the parameter $n$ which  is related to the dumping of the feature, and larger values are associated to a steeper change of the potential, as shown in Figs. \ref{vnback}-\ref{nback}. The slow-roll parameters show an oscillation around the feature time $t_0$ with a larger amplitude for larger $n$, since a steeper potential change is also associated to larger derivatives of the Hubble parameter (see Ref.~\cite{GallegoCadavid:2015hld}.) We will be particularly interested in the effect of the parameter $n$ which will give a unique characteristic on the abundance of PBH, as we will see below.

In this paper we study the abundance of PBH seeded by peaks of the primordial curvature perturbations spectrum of models with a local feature. The tuning of the parameters of the features allows to produce PBHs in different mass ranges.

\section{Curvature perturbations}
The study of curvature perturbations is attained by expanding perturbatively the action with respect to the background FRLW solution. In the comoving gauge the second order action for scalar perturbations is \cite{Maldacena:2003}
\be\label{eq:s2}
 S_2 = \int dt d^3x\left[a^3 \epsilon \dot{\mathcal{R}_{c}^2}-a\epsilon(\partial \mathcal{R}_{c})^2 \right] \,.
\ee
Taking the Fourier transform of the Euler-Lagrange equations of the previous action and using the conformal time $d\tau \equiv dt/a$ we obtain the equation of motion for the primordial curvature perturbation
\begin{equation}\label{eq:cpe}
  \mathcal{R}_{c}''(k) + 2 \frac{z'}{z} \mathcal{R}_{c}'(k) + k^2 \mathcal{R}_{c}(k) = 0,
\end{equation}
where $z\equiv a\sqrt{2 \epsilon}$~\cite{Langlois:2010xc}, $k$ is the comoving wave number, and primes denote derivatives with respect to conformal time. We take the Bunch-Davies vacuum \cite{Bunch:1978yq, Noumi:2014zqa, Langlois:2010xc}
\be \label{eq:ic1}
\mathcal{R}_c(\tau,k) = \frac{v(\tau,k)}{z(\tau)} ,
\ee
as the initial conditions for \eqn{eq:cpe}, where
\begin{equation}
  v(\tau,k)=\frac{e^{-\mathrm{i}  k  \tau}}{\sqrt{2 k }}\left(1-\frac{\mathrm{i}}{ k \tau}\right)\,.
\end{equation}

\section{Super-horizon growth of $\R_c$ in single field models}
Following \cite{Romano:2016gop}, we will briefly review the mechanism by which curvature perturbations can grow on super horizon scales in single scalar field models.
The equation for the curvature perturbation on comoving slices can be written as
\be
\frac{\partial}{\partial t}\left(\frac{a^3\eps}{c_s^2}
\frac{\partial}{\partial t}\R_c\right)-a\eps \Delta \R_c=0 \label{eomR2} \,,
\ee
and  after re-writing the time derivative in terms 
of the derivative respect to the scale factor $a$, we can obtain the solution on super-horizon 
scales 
\ba
\R_c &\propto& \int^a\frac{da}{a} f(a)\,;
\quad 
f(a)\equiv{\frac{c_s^2(a)}{Ha^3 \epsilon(a)}} \,. \label{Rca}
\ea
%
%where we have introduced the function $f(a)$ for later convenience.
In  slow-roll inflation $c_s^2$ and $\epsilon$ are
 slowly varying, so that the integrand is asymptotically constant,
implying the $\R_c$ conservation. 

%The necessary and sufficient condition for super-horizon freezing is that there exists some $\delta>0$ for which
%\be
%\lim _{a\to\infty}a^\delta f(a)=0 . \label{freezeF}
%\ee
Since $\epsilon=-\dot H/H^2\ll1$,  we can neglect the time dependence of
$H$ in \eqn{Rca}, while $\epsilon$ and $c_s^2$ 
may still vary rapidly in time. 
Assuming $\epsilon\approx a^{-n}$ and $c_s^2\approx a^q$ we get 
\be
f \propto a^{q+n-3} \,,
\ee
from which the  condition for freezing is 
\be
q+n-3<0 \,.
\ee 
If the condition is violated, i.e. $q+n-3\geq 0$, 
 the solution (\ref{Rca}) can grow on super-horizon scales. In single filed models with minimal gravity coupling $q=0$, which implied that $\epsilon$ should decrease sufficiently rapidly. This is exactly what happens in the models we study in this paper, where the feature of the potential induces a temporary violation of the slow-roll regime, and a fast decrease of $\epsilon$, with $n>3$.

\subsection{Primordial spectrum of curvature perturbations}\label{pscp}
The two-point correlation function of primordial curvature perturbations is given
by~\cite{Ade:2015lrj,Planck:2013jfk}
\begin{equation}
 \Braket{ \hat{\mathcal{R}}_c(\vec{k}_1) \hat{\mathcal{R}}_c(\vec{k}_2) } \equiv (2\pi)^3
\frac{2\pi^2}{k^3} P_{\mathcal{R}_c}(k) \delta^{(3)}(\vec{k}_1+\vec{k}_2) \, ,
\end{equation}
where the modes are evaluated after horizon crossing time and the power spectrum of primordial curvature perturbations is defined as
\be
P_{\mathcal{R}_{c}}(k) \equiv \frac{2k^3}{(2\pi)^2}|\mathcal{R}_{c}(k)|^2.
\ee

We define $k_0$ as the scale exiting the horizon at the feature time $t_0$, $k_0=-1/\tau_0$, where $\tau_0$ is the value of conformal time corresponding to $t_0$.  Oscillations occur around $k_0$, and their location can be controlled by changing $\phi_0$. 

\begin{figure}
 \begin{minipage}{.45\textwidth}
  \includegraphics[scale=1]{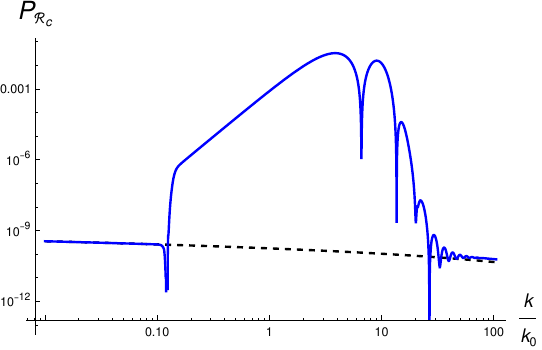}
  \end{minipage}
  \caption{The numerically computed $P_{\R}$ is plotted for $\lambda=1.8196427\times 10^{-11}$, $\sigma=0.02$, and $n=1$. The dashed black lines correspond to the featureless behavior.}
\label{Pplot}
\end{figure}

A representative power spectrum behavior able to produce a significant amount of PBH around the scale $k_0$ is shown in Fig. \ref{Pplot}. There it can seen the oscillations of the power spectrum around the featured scale.
%**************************************************
%**************************************************
\section{Production and abundance of PBH}
In this section we would briefly review the production and abundance of PBH. We will closely follow Ref.~\cite{Ozsoy:2023ryl}. 

The characteristic mass of PBH can be related to the mass contained within the Hubble horizon at the formation time $H^{-1}_{(\rm f)}$ given by \cite{Carr:1975qj}
\be\label{pbhm}
M^{(\rm f)}_{\rm pbh} = \gamma\, M_H ^{(\rm f)} \, ,
\ee
where $\gamma$ is an efficiency factor,  the sub/super scripts ``\textrm{f}" denote quantities evaluated at the time of PBH formation, and
\be
M_H (t) \equiv \frac{4\pi}{3} \rho(t) H(t)^{-3} ,
\ee
is the time-dependent horizon mass.

The PBH mass at formation time can be related to the characteristic size of the perturbations that leave the horizon during inflation by
\be \label{pbhmk}
M^{(\rm f)}_{\rm pbh}(k_{\rm pbh}) \simeq \left(\frac{\gamma}{0.2}\right)\left(\frac{g_{*}\left(T_{\rm f}\right)}{106.75}\right)^{-1 / 6}\left(\frac{k_{\rm pbh}}{1.753 \times 10^{13}\, \mathrm{Mpc}^{-1}}\right)^{-2}\, 10^{-14}\, M_{\odot}\,.
\ee
where we have assumed that the  effective number of relativistic degrees of freedom in energy density and entropy are equal \cite{Baumann_2022, Ozsoy:2023ryl}.

The present-day  PBH fraction $f_{\rm pbh}$ is defined as the ratio between the energy density of dark matter and PBHs, and  can be written in terms of the PBH mass $M^{(\rm f)}_{\rm pbh}$ and the abundance at formation $\beta$ by \cite{Ozsoy:2023ryl}
\be\label{betaf}
f_{\rm pbh} \left(M^{(\rm f)}_{\rm pbh}\right) \simeq 7.52 \times 10^{15}  \left(\frac{\gamma}{0.2}\right)^{1/2} \left(\frac{g_{*}\left(T_{\rm f}\right)}{106.75}\right)^{1 / 12} \left(\frac{M^{(\rm f)}_{\rm pbh}}{ 10^{-14} M_{\odot}}\right)^{-1/2}\, \beta \left(M^{(\rm f)}_{\rm pbh}\right) \,.
\ee

The PBH abundance at formation can be interpreted as the fraction $\beta$ of local regions in the universe that has a density larger than a certain threshold $\delta_c$. The simplest formalism for estimating this abundance is the so-called Press-Schechter model of gravitational collapse \cite{Press:1973iz}, where $\beta$ is given by
\be\label{beta}
\beta \equiv \frac{\rho_{_\mathrm{{pbh}}}}{\rho} \bigg|_{a = a_{\rm f}} = \int_{\delta_{\rm c}}^{\infty}P(\delta)\,\, \d \delta = \int_{\delta_{\rm c}}^{\infty} \frac{\mathrm{d} \delta}{\sqrt{2 \pi} \sigma_{\rm pbh}} \exp \left(-\frac{\delta^{2}}{2 \sigma^{2}_{\rm pbh}}\right),
\ee
where $P(\delta)$ is the probability distribution function (PDF), describing the probability that a given fluctuation have an overdensity $\delta$. In the last equality we assume that the density perturbations follow a Gaussian distribution with zero mean and variance $\sigma_{\rm pbh}$. Moreover we explicitly assume that a perturbation will collapse to form black holes when its amplitude is larger than a critical value $\delta_c$.

%***************************************************
Assuming $\delta_{\rm c} \gg \sigma_{\rm pbh}$, which we show below is a good approximation, \eqn{beta} can be written as
\be\label{betaG}
\beta \simeq \frac{1}{2} \rm{Erfc}\left(\frac{\delta_{\rm c}}{\sqrt{2}\sigma_{\rm pbh}}\right) \simeq \frac{ \sigma_{\rm pbh}}{\sqrt{2 \pi} \delta_{\rm c} } \exp \left(-\frac{\delta_{\rm c} ^{2}}{2 \sigma^2_{\rm pbh}}\right),
\ee
where ${\rm Erfc}(x)$ is the complementary error function. Notice from \eqn{betaG} that the PBH abundance at the time of formation is exponentially sensitive to the variance of the distribution. Therefore, even a small change in the values of the model parameters, which might only slightly affect the primordial power spectrum, can have a substantial impact on the resulting PBH abundance. This exponential sensitivity underscores the critical importance of accurately determining the model parameters. These results are compatible with those of Refs. \cite{Cole:2023wyx, Mishra:2019pzq}.

We now discuss the relation between the abundance of PBH and the amplitude of the comoving curvature perturbations generated during  inflation.
%**************************************************
\begin{figure}
 \begin{minipage}{.45\textwidth}
  \includegraphics[scale=1]{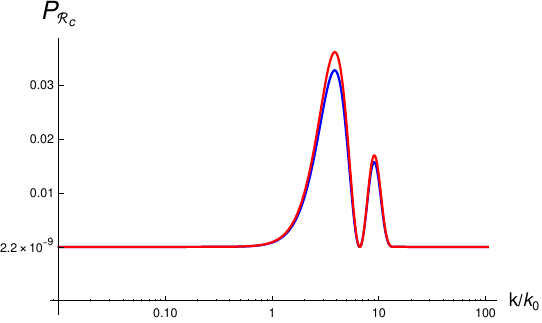}
  \end{minipage}
  \caption{The numerically computed $P_{\R}$ is  plotted for $\sigma=0.02$, $n=1$,  $\lambda=1.8196427\times10^{-11}$ (blue) and $\lambda=1.81964272\times10^{-11}$ (red). We use a linear scale for the vertical axis to show the difference between the two spectra. %The typical oscillations of the primordial spectrum are not seen because we choose a linear vertical axis in order to show the relative amplitude of the spectra.
  }
\label{lambdapert}
\end{figure}

\begin{figure}
 \begin{minipage}{.45\textwidth}
  \includegraphics[scale=1]{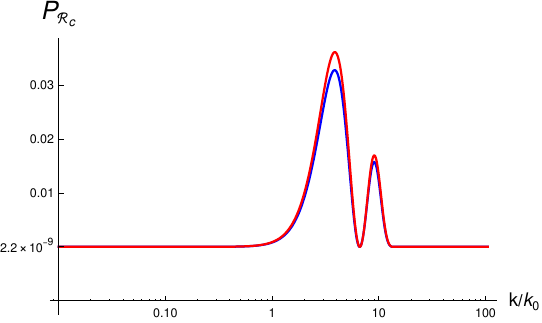}
  \end{minipage}
 \caption{The numerically computed $P_{\R}$ is  plotted for $\lambda=1.8196427\times10^{-11}$, $n=1$,  $\sigma=0.02$ (blue) and $\sigma=0.0199998$ (red). We use a linear scale for the vertical axis to show the difference between the two spectra. }
\label{sigmapert}
\end{figure}

\begin{figure}
 \begin{minipage}{.45\textwidth}
  \includegraphics[scale=1]{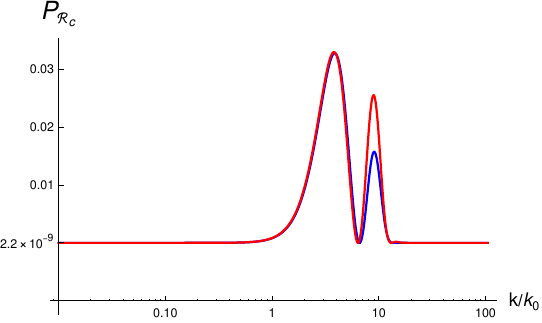}
  \end{minipage}
 \caption{The numerically computed $P_{\R}$ is  plotted for
 $\{\lambda=1.8196427\times10^{-11}, \sigma=0.02, n=1\}$ (blue)  $\lambda=1.707869\times10^{-11}$, and $\{\sigma=0.02,n=2\}$ (red). Notice we have to decrease the value of the $\lambda$ parameter when choosing $n=2$, or the power spectrum would be too large, giving $f_{\rm pbh}>1$. Notice also that the second peak of the red line is much larger than the second peak of the blue line. We use a linear scale for the vertical axis to show the difference between the two spectra.}
\label{npert}
\end{figure}
%**************************************************
\subsection{The comoving curvature perturbations and the abundance of PBH}\label{S2p3}

The power spectrum of over-density is related to the comoving curvature perturbation by
\be\label{dvsR}
\mathcal{P}_\delta (k) \simeq \frac{16}{81} \left(\frac{k}{a H}\right)^4 \mathcal{P}_{\mathcal{R}_c} (k).
\ee

The variance of density contrast can be related to the primordial power spectrum as \cite{Young:2014ana,Sasaki:2018dmp}
\be\label{vardef}
\sigma^2_{\rm pbh} (R) \equiv \langle \delta^2 \rangle_{R} = \int_{0}^{\infty} \d \ln q\,\, \mathcal{W}^2(q, R)\, \mathcal{P}_{\delta} (q) \simeq \frac{16}{81} \int_{0}^{\infty} \d \ln q\,\, \mathcal{W}^2(q, R)\, (qR)^4\,\, \mathcal{P}_{\mathcal{R}}(q)
\ee
where $\mathcal{W}$ is a window function smoothing $\delta$ over the comoving scale $R \approx k^{-1} \approx (aH)^{-1}$. In this paper we will use the volume-normalized Gaussian window function, whose Fourier transform is given by
\be
\mathcal{W}(k, R) = \exp\left({-\frac{k^2 R^2}{2}}\right)\,.
\ee

In Figs. \ref{lambdapert}-\ref{npert} we can see the power spectra used to calculate the variance $\sigma_{\rm pbh}$. This variance is later use to calculate $f_{\rm pbh}$ as a function of the mass at the formation time. Notice the small changes in the values of the model parameters,  resulting in minor variations in the power spectrum. We will see below that these changes have a significant impact on the PBH abundance. This is because, as previously mentioned, the PBH abundance is exponentially sensitive to the variance of the density distribution.

\begin{figure}[h]
 \centering
 \includegraphics[scale=1]{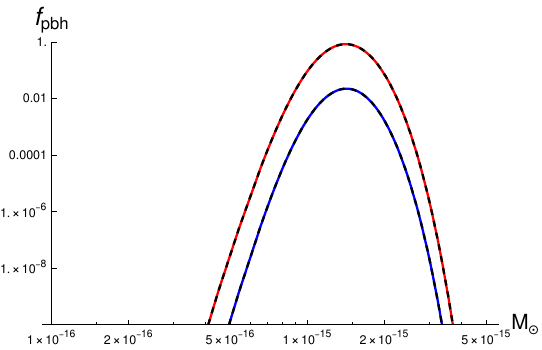}
 % fpbhplot_m1_123.pdf: 0x0 px, 300dpi, 0.00x0.00 cm, bb=
 \caption{The numerically computed $f_{\rm pbh}$ is  plotted for $\sigma=0.02$, $n=1$, $\lambda=1.8196427\times10^{-11}$ (blue) and $\lambda=1.81964272\times10^{-11}$ (red). The dashed black lines are the analytical approximations given by \eqn{betaG}.}
 \label{fig:fpbhm1}
\end{figure}

\begin{figure}[h]
 \centering
 \includegraphics[scale=1]{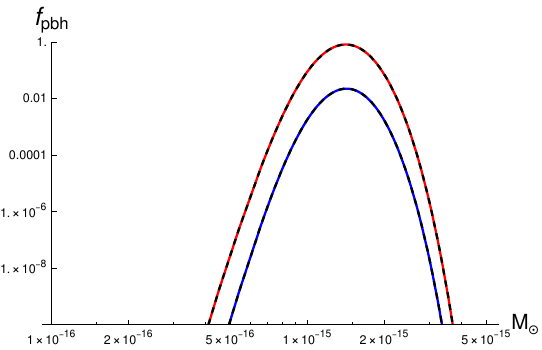}
 \caption{The numerically computed $f_{\rm pbh}$ is  plotted for $\lambda=1.8196427\times10^{-11}$, $n=1$,  $\sigma=0.02$ (blue) and $\sigma=0.0199998$ (red). The dashed black lines are the analytical approximations given by \eqn{betaG}.}
 \label{fig:fpbhm2}
\end{figure}

\begin{figure}[h]
 \centering
 \includegraphics[scale=1]{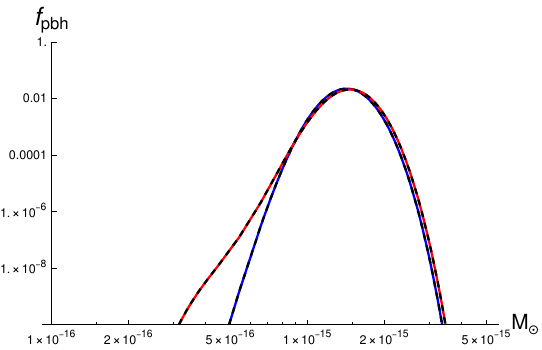}
 % fpbhplot_m1_123.pdf: 0x0 px, 300dpi, 0.00x0.00 cm, bb=
 \caption{The numerically computed $f_{\rm pbh}$ is  plotted for 
 $\{\lambda=1.8196427\times10^{-11}, \sigma=0.02, n=1\}$ (blue) and 
 $\{\lambda=1.707869\times10^{-11}, \sigma=0.02,  n=2\}$ (red). The dashed black lines are the analytical approximations given by \eqn{betaG}. Notice that in this case we have to decrease the value of the $\lambda$ parameter when choosing $n=2$, or  $f_{\rm pbh} \gg 1$. Notice also that, due to the second peak in Fig. \ref{npert}, $f_{\rm pbh}$ is not symmetric.}
 \label{fig:fpbhm3}
\end{figure}

\iffalse
\begin{figure}[h]
 \centering
 \includegraphics[scale=1]{different_n.pdf}
 \caption{The quantity $f_{\rm pbh}$ is plotted as a function of the mass for different values of the parameter $n$.}
 \label{fig:differentn}
\end{figure}
\fi

\begin{figure}[h]
 \centering
 \includegraphics[scale=1]{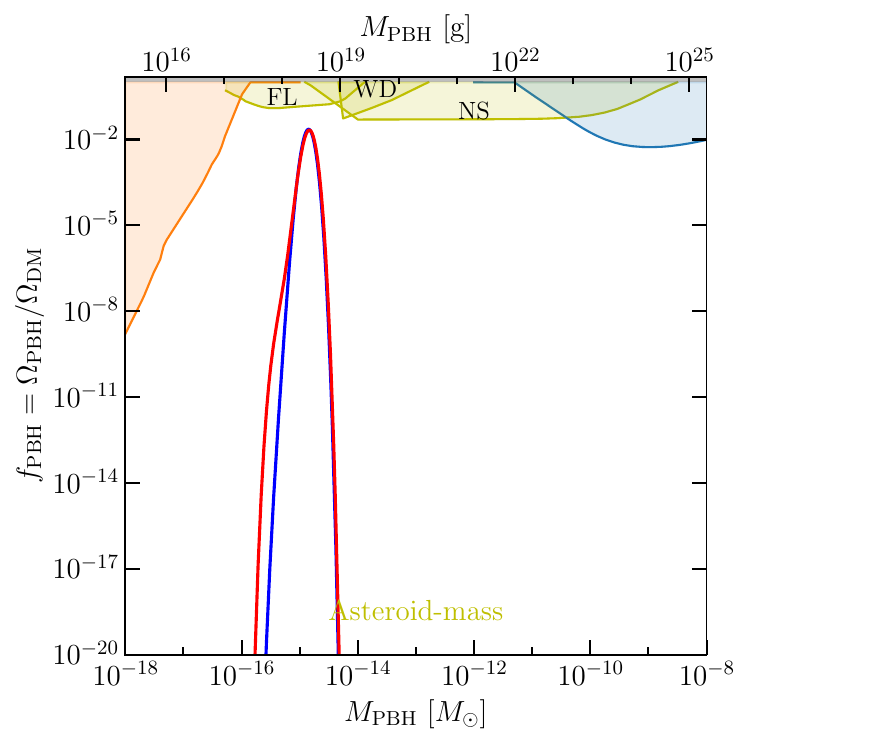}
 \caption{Asteroid mass constraints on $f_{\rm pbh}$ is  plotted for $\lambda=1.8196427\times10^{-11}$, $\sigma=0.02$, and $n=1$ (blue) and $\lambda=1.707869\times10^{-11}$, $\sigma=0.02$, and $n=2$ (red). % using eqs. \ref{betaf} and \ref{beta}. 
 Notice that in this case we have to decrease the value of the $\lambda$ parameter when choosing $n=2$, or  $f_{\rm pbh} \gg 1$. Notice also that, due to the second peak in Fig. \ref{npert}, $f_{\rm pbh}$ is not symmetric.}
 \label{fig:constraints}
\end{figure}

\section{Production of asteroid mass PBHs}

As mentioned before, the parameter space of our model allows for the production of PBH of different masses, ranging from asteroid masses to hundreds of solar masses. As can be seen in \eqn{pbhmk} the PBH mass at formation time is related to the scale of the perturbation at horizon exit. Thus we can tune the parameter $k_0$, or equivalently $\phi_0$, to produce PBH of the desired mass. As an example of the general mechanism, in this section we will consider features producing PBHs in the the asteroid mass range (approximately $10^{-17} \lesssim M_{\mathrm{PBH}} \left[M_{\odot}\right] \lesssim 10^{-12}$). The results are shown in Figs. \ref{fig:fpbhm1}-\ref{fig:fpbhm3}. In this case the range of  scales associated with PBH formation is $10^{12} \lesssim k_{\rm pbh} \,[ \mathrm{Mpc}^{-1}]\, \lesssim 10^{15}$ and we  will set the feature scale at $k_0=1.753\times 10^{13}$ Mpc${}^{-1}$. In Figs. \ref{fig:fpbhm1}-\ref{fig:fpbhm3} it can be seen that \eqn{betaG} is a good approximation to calculate the abundance at formation $\beta$.

The asteroid mass range is important for several reasons \cite{Green_2021, Carr:2020gox}. Asteroid-mass PBHs can be detected through microlensing events, where the PBH’s gravitational field magnifies the light from a background star. Surveys like the Optical Gravitational Lensing Experiment (OGLE) can detect PBHs in this mass range, providing constraints on their abundance. These PBHs are potential dark matter candidates, exhibiting unique dynamical and observational signatures. Furthermore, asteroid-mass PBHs could produce a stochastic gravitational wave background detectable by instruments like the Laser Interferometer Space Antenna (LISA), offering insights into their abundance and distribution.

In Fig. \ref{fig:constraints} we show the current asteroid mass constraints on $f_{\rm pbh}$ \cite{bradley_j_kavanagh_2019_3538999}. The contraints shown are the Femto-lensing (FL) \cite{Barnacka_2012}, White Dwarf explosions (WD) \cite{Graham:2015apa}, and the Neutron star disruption (NS) \cite{Capela:2013yf} contraints. The orange and blue shaded areas in the figure correspond to the evaporation and microlensing constraints, respectively \cite{Green_2021, Carr:2020gox}.

\section{Conclusions}
We have studied the effects of local features of the inflation potential  on the abundance of PBHs. 
These features only modify the potential in a limited range of the scalar field values, and consequently only affect the spectrum in a narrow range of scales, which leave the horizon during the time interval corresponding to the modification of the potential.
We have studied the effects of the different parameters of the feature on PBH abundance, showing how they provide a flexible and general mechanism able to produce PBHs in different mass ranges.
We have then given a specific example of a model producing PBHs with abundance compatible with the asteroid mass constraints, confirming previous results \cite{Cole:2023wyx} about high sensitivity of the PBH abundance to the model parameters.

In this paper we have assumed that density perturbations follow a Gaussian distribution, but in presence of features non gaussianity of the primordial perturbations can be important, and this could have important effects on the PBH abundance calculation \cite{Germani:2018jgr}. In the future it will be interesting to compute the non-gaussian contributions to the PBH abundance in the local features models we have considered.

%\newpage
\acknowledgments
A.G.C. was funded by Agencia Nacional de Investigación y Desarrollo  ANID through the FONDECYT postdoctoral Grant No. 3210512.

\bibliography{Bibliography,BibliographyPBH}
\bibliographystyle{h-physrev4}
\end{document}